\documentclass[preprint,showpacs,amsmath,amssymb]{revtex4}
\usepackage{graphicx,amsmath,amssymb,bbm,times}
\usepackage{epsfig}
\begin{document}
\title{Photon statistics in the macroscopic realm measured without photon-counters. }
\author{Alessandra Andreoni}\email{alessandra.andreoni@uninsubria.it}
\affiliation{Dipartimento di Fisica e Matematica, Universit\`a degli Studi dell'Insubria and C.N.I.S.M., U.d.R. Como, I-22100, Como, Italy}
\author{Maria Bondani}
\affiliation{National Laboratory for Ultrafast and Ultraintense
Optical Science - C.N.R.-I.N.F.M. and C.N.I.S.M., U.d.R. Como, I-22100, Como, Italy}
\date{\today}
\begin{abstract}
In a macroscopic realm, in which photons are too many for being counted by any photon counting detector, photon statistics can be measured by using detectors simply endowed with linear response. We insert one of such detectors in a conventional photon-counting apparatus, which returns a voltage every time the detector responds to light by generating a number of elementary charges via its primary photo-detection process. We only assume that, when a single charge is photo-generated, the probability density of the voltages is a distribution that is narrow with respect to its mean value. Under this hypothesis the output voltages can be suitably binned so that their probability distribution is the same as that of the photo-generated charges, that is, of the detected photons.
\end{abstract}
\maketitle
\section{Introduction}

Measuring photon statistics is a useful approach to understand the behavior of any system that includes electromagnetic radiation as a part. The investigation of such systems may pertain to physics, from astronomy to physics of the matter \cite{Wolf}, as well as to other natural sciences, for instance biology \cite{Wang}. The availability of photon-counting detectors and methods suitable for any situation as to spectral and intensity characteristics of the light to be measured would then be extremely desirable. The coverage of the most different spectral ranges is a goal that is pursued by the search of novel primary photo-detection processes, including thermal processes occurring at cryogenic temperatures. Among the detectors that operate, in essence, as microcalorimeters we mention a superconducting transition-edge sensor (TES) with tungsten as the active device material that was recently demonstrated to work as a photon-counter endowed with almost unitary quantum efficiency from UV-vis to telecom wavelengths \cite{Lita}. However, we recognize that detectors based on quantum interactions between photons and sensitive material are largely more used than thermal detectors for measuring photon statistics. Detectors based on either external primary processes (\textit{e.g.} electron photo-emission) or internal primary processes (\textit{e.g.} photo-generation of carriers by either photo-voltaic or photo-conductive effects) ensure reasonable values of the detection quantum efficiency, $\eta_\mathrm{q}$ in the visible and near-IR spectral ranges. The main difficulty that still remains with these detectors is that of measuring photon statistics when the charges photo-generated in the samples are too many to be counted. Among photoemissive detectors only few produce distinct outputs when the number of photo-electrons, $m$, changes by a unit. The best ones are photomultiplier tubes (PMT's) \cite{noiRSI, noiJMO} and hybrid photo-detectors (HPD's) \cite{noiJMO} that can count up to  $m\approx 5$. Relatively more numerous are the photo-emissive detectors that are endowed with sufficiently high and sharp gain to provide a sizeable charge in the anodic pulse output for $m\geq1$  definitely distinguished from that for $m=0$. They are PMT's available since the 50's that were used for the first measurements of light statistics \cite{Arecchi, Freed, Martiens, Smith}. Nowadays single-photon detectors exist that are based on the most different primary photo-detection processes and offer a remedy to the lack of good photon counters. In fact, the light to be measured can be split either in space or in time prior to detection so that at most one photon at a time hits the detector sensitive area. However it must be recognized that these techniques invented for counting photons with intensified CCD cameras \cite{iCCD} and multi-pixel and/or position sensitive single-photon detectors \cite{Kok,SiPMT,Jiang, Yama} (spatial splitting) or single-photon avalanche photodiodes (temporal splitting) \cite{Banas,SPAD} are rather cumbersome. Their adoption is only justified by the impossibility of performing direct measurements with photon counters when the number of detected photons becomes macroscopic.

The work described here concerns the direct measurement of the detected-photon statistical distribution, $P_m$, and is motivated by the fact that, in many of the systems for which measuring photon statistics is relevant, artificially lowering $m$ is not permitted either by attenuating the light or by shortening the measuring time, $T_\mathrm{M}$. This is the case of fields that modify their properties upon attenuation and, obviously, of pulse fields in which $T_\mathrm{M}$ cannot be shorter than the light pulse duration. It is worth noting that measuring photon statistics when $m\geq1$ in $T_\mathrm{M}$ is a problem that has been faced since the 60's. In particular Arecchi \emph{et al.} \cite{IEEE} suggested a "linear method" in which the PMT anodic charge corresponding to the photons detected in $T_\mathrm{M}$ was recorded. Moreover these authors demonstrated that calculating the moments of the statistical distribution of this charge and those of the single-electron response (SER) distribution allows obtaining the moments of $P_m$. Such a result has been used to verify the agreement with the theoretical $P_m$ moments up to second order \cite{ArecchiDegiorgio}. However, using it to recover $P_m$ would be at least cumbersome owing to the need of accurate evaluations of SER distribution moments at any order. We will show that any detector based on either an internal or an external primary photo-detection process and endowed with two properties rather commonly encountered, allow measuring $P_m$ in a macroscopic realm in which photon-counters do not exist. The two properties are: (i) the detector response must be linear up to the maximum $m$ of the measurement; (ii) the response for $m=1$ must produce a standard deviation of the output data that is sufficiently smaller than the mean value. We further specify that the detector can be endowed with an internal gain.

\section{Model}
\begin{figure}[h]
\begin{center}{
 \includegraphics[angle=270, width=0.45\textwidth]{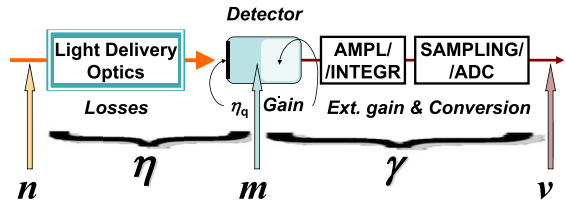} }
\caption{(Color online) Measuring apparatus.} \label{fig:1}
\end{center}
\end{figure}

With the help of Fig.~\ref{fig:1} we first examine how the detector output is processed in a typical direct statistical measurement. Normally it is amplified and integrated over a temporal gate, whose duration fixes the value of the measure time $T_\mathrm{M}$ when a continuous wave light is to be measured. In the case of pulsed light, the gate is synchronous and covers the  $T_\mathrm{M}$ interval in
which the current output pulse of the detector occurs. The signal is sampled and digitized afterward and converted to a voltage $v$.%
As indicated in the figure, we represent the overall $m$-to-$v$ conversion by a single factor, $\gamma$. Here we will explicitly take into account the statistical distribution of the probability density $p_\gamma$ of this conversion factor.  The left-hand side of Fig.~\ref{fig:1} illustrates the link between the probability density of detecting $m$ photons, $P_m$, and that, $P_n$, of having $n$ photons in the field. Considering the effects of the optics that delivers the light to the detector and of the detector quantum efficiency $\eta_\mathrm{q}<1$ and representing these concomitant losses by an overall photon-detection efficiency, $\eta$, lead to \cite{MandelandWolf}
\begin{equation}
 P_m=\sum_{n=m}^{+\infty} \left(
 \begin{array}{c}n\\m\end{array}\right)
 \eta^m (1-\eta)^{n-m} P_n\ .\label{eq:phel}
\end{equation}
Obviously in any experiment the value of $\eta$ is up-limited by the product of $\eta_\mathrm{q}$  times the coupling efficiency of the optical delivery system, but can be diminished at will if filters are inserted into the system that delivers the light to the detector. We point out that, as we deal with direct statistical measurements, neither the delivery optics includes a fiber looping beam splitter nor the detector is a position sensitive one.

Our aim is to recover $P_m$ for an arbitrary $P_n$ starting from the only experimental data available, which are the $v$ voltage values recorded for an ensemble of measurements performed with given $\eta$ by using an apparatus characterized by a conversion factor $\gamma$ with probability distribution $p_\gamma$, mean value $\bar{\gamma}$ and variance $\sigma^2$. In the following we indicate by $P_v$ the probability density of the $v$  variable. We assess that we can "measure" $P_v$ as the distribution that we would obtain by casting the experimental $v$ values of an ensemble of measurements into a histogram normalized to its integral. For ease of writing we represent the bin width by $dv$, though the variable $v$ is our digitized output. The zero of the $v$ scale of the "measured" $P_v$ is set to be equal to the mean value of the distribution recorded in a separate experiment performed in the absence of light.
As the events of having different values of detected photons (i.e. elementary charges generated by the primary photo detection process) are mutually exclusive we can write
\begin{eqnarray}
 P_v&=&P_{m=0}\mathrm{P}_v^{(0)}+P_{m=1}\mathrm{P}_v^{(1)}+P_{m=2}\mathrm{P}_v^{(2)}+\ldots\nonumber\\
 &=&\sum_{k=0}^{+\infty} P_{m=k}\mathrm{P}_v^{(k)}\ ,\label{eq:Pv}
\end{eqnarray}
in which $\mathrm{P}_v^{(k)}$ is defined as the probability density of the voltage values $v^{(k)}$ that are recorded in the events with $k$ detected photons. In Eq.~(\ref{eq:Pv}), $\mathrm{P}_v^{(0)}$  is the probability distribution measured in the absence of light, for which we remind that $\int_{-\infty}^{\infty}v\mathrm{P}_v^{(0)}dv=0$. We note that $P_v$ obviously reproduces $P_m$ if the measuring apparatus has photon-counting capability \cite{noiRSI, noiJMO} whereas, when the $\mathrm{P}_v^{(k)}$'s do not lead to separate peaks in $P_v$, Eq.~(\ref{eq:Pv}) seems to be useless to reconstruct $P_m$. The latter is exactly the case examined in this paper.

We consider the central moments $\mu_r(v)=\langle(v-\langle v\rangle)^r\rangle$ corresponding to the experimental $P_v$  and try to relate them to the $\mu_r(m)=\langle(m-\langle m\rangle)^r\rangle$  central moments corresponding to the unknown $P_m$ probability density. By using properties of the $\mathrm{P}_v^{(k)}$ distributions to express  $\mu_r(v)$, we will find relations to  $\mu_r(m)$ that provide a method to reconstruct $P_m$. As $\mathrm{P}_v^{(1)}$  can be identified with the probability distribution $p_\gamma$ of the conversion factor $\gamma$, we obviously have $v^{(1)}=\gamma$. Owing to the hypothesis that the detector response is linear, detecting $k>1$ elementary charges corresponds to the occurrence of independent events, thus $v^{(k)}=\sum_{i=1}^k\gamma_i$, in which all $\gamma_i$  are distributed according to $p_\gamma$, and  $\mathrm{P}_v^{(k)}=\mathrm{P}_v^{(1)}\ast\mathrm{P}_v^{(1)}\ast\ldots\ast\mathrm{P}_v^{(1)}$ for $k$  times. Thus we can exploit the following property of the cumulants \cite{MandelandWolf}:
\begin{eqnarray}
 \kappa_r^{(\sum_{i=1}^k \gamma_i)}=\sum_{i=1}^k \kappa_r^{(\gamma_i)}\ .\label{eq:cumul}
\end{eqnarray}
By reminding that the lowest order cumulants are $\kappa_1^{(x)}= \langle x\rangle$, $\kappa_2^{(x)}= \mu_2(x)$, $\kappa_3^{(x)}= \mu_3(x)$, $\kappa_4^{(x)}= \mu_4(x)-3[\mu_2(x)]^2$, and $\kappa_5^{(x)}= \mu_5(x)-10\mu_2(x)\mu_3(x)$, for the cumulants of the conversion/amplification factor we find: $\kappa_1^{(\gamma_i)}= \bar{\gamma}$, $\kappa_2^{(\gamma_i)}=\sigma^2$, $\kappa_3^{(\gamma_i)}= \tilde{\mu}_3$, $\kappa_4^{(\gamma_i)}= \tilde{\mu}_4 -3\sigma^4$, $\kappa_5^{(\gamma_i)}= \tilde{\mu}_5-10\sigma^2\tilde{\mu}_3$, being $\tilde{\mu}_r$ the values assumed for the central moments of $p_\gamma$.

We start by using Eq.~(\ref{eq:cumul}) with $r=1$ and then Eq.~(\ref{eq:Pv}) to calculate the mean value of $v$:
\begin{eqnarray}
 \langle v\rangle = \bar{\gamma}\sum_{k=0}^{+\infty} k P_{m=k}= \langle m\rangle\bar{\gamma}\ .\label{eq:vAVER}
\end{eqnarray}
We now calculate the $\mu_r(v)$ moments by applying Eq.~(\ref{eq:Pv}):
\begin{eqnarray}
 \mu_r(v)&=&\sum_{k=0}^{+\infty} P_{m=k}\int_{-\infty}^{+\infty}(v-\langle v\rangle)^r \mathrm{P}_v^{(k)}dv\nonumber\\
 &=&\sum_{k=0}^{+\infty} P_{m=k} \mu_r(v^{(k)})\ ,\label{eq:muV}
\end{eqnarray}
where we used $\mu_r(v^{(k)})\equiv\int_{-\infty}^{+\infty}(v-\langle v\rangle)^r \mathrm{P}_v^{(k)}dv$.

For $r=1$ Eq.~(\ref{eq:muV}) obviously vanishes. For $r\geq 2$ we make use of the binomial expansion
\begin{eqnarray}
 (v-\langle v\rangle)^r = \sum_{j=0}^{r} \left(
 \begin{array}{c}r\\j\end{array}\right)
 v^j (-\langle v\rangle)^{r-j}\ ,\label{eq:binom}
\end{eqnarray}
which, once substituted into Eq.~(\ref{eq:muV}) and using Eq.~(\ref{eq:vAVER}), gives
\begin{eqnarray}
 \mu_r(v)=\sum_{j=0}^{r} \left(
 \begin{array}{c}r\\j\end{array}\right)
 (-\langle m\rangle \bar{\gamma})^{r-j}
 \sum_{k=0}^{+\infty} P_{m=k}\mu'_j(v^{(k)})\ .\label{eq:muV2}
\end{eqnarray}
In Eq.~(\ref{eq:muV2}) the "prime" distinguishes the moments from the central moments. The recursion formula that relates the moments to the cumulants \cite{refCUM} in our case reads
\begin{eqnarray}
 \mu'_j(v^{(k)})= \kappa_j^{(v^{(k)})}+\sum_{s=1}^{j-1} \left(
 \begin{array}{c}j-1\\s-1\end{array}\right)
 \kappa_s^{(v^{(k)})}\mu'_{j-s}(v^{(k)})\ ,\label{eq:muVprime}
\end{eqnarray}
from which it can be shown that the $j$-th order moment, $ \mu'_j(v^{(k)})$, is a polynomial of the first $j$ cumulants, $\kappa_s^{(v^{(k)})}$  with $s=1,2,\ldots,j$. Thus in Eq.~(\ref{eq:muVprime}):
\begin{eqnarray}
 \mu'_1(v^{(k)})&=& \kappa_1^{(v^{(k)})}\nonumber\\
 \mu'_2(v^{(k)})&=& \kappa_2^{(v^{(k)})}+(\kappa_1^{(v^{(k)})})^2\nonumber\\
 \mu'_3(v^{(k)})&=& \kappa_3^{(v^{(k)})}+3\kappa_2^{(v^{(k)})}\kappa_1^{(v^{(k)})} +(\kappa_1^{(v^{(k)})})^3\nonumber\\
 \mu'_4(v^{(k)})&=& \kappa_4^{(v^{(k)})}+4\kappa_3^{(v^{(k)})}\kappa_1^{(v^{(k)})} +3(\kappa_2^{(v^{(k)})})^2\nonumber\\
 &&+6\kappa_2^{(v^{(k)})}(\kappa_1^{(v^{(k)})})^2+(\kappa_1^{(v^{(k)})})^4\nonumber\\
 \mu'_5(v^{(k)})&=& \kappa_5^{(v^{(k)})}+5\kappa_4^{(v^{(k)})}\kappa_1^{(v^{(k)})}
 +10\kappa_3^{(v^{(k)})}\kappa_2^{(v^{(k)})}\nonumber\\
 &&+10\kappa_3^{(v^{(k)})}(\kappa_1^{(v^{(k)})})^2 +15(\kappa_2^{(v^{(k)})})^2\kappa_1^{(v^{(k)})}\nonumber\\
 &&+10\kappa_2^{(v^{(k)})}(\kappa_1^{(v^{(k)})})^3
 +(\kappa_1^{(v^{(k)})})^5\nonumber\\
 \mu'_6(v^{(k)})&=&\ldots\nonumber\ ,
\end{eqnarray}
where the coefficients of the different terms are those that occur in the Fa\`a di Bruno's formula. By using Eq.~(\ref{eq:cumul}) we can rewrite these terms in the form
\begin{eqnarray}
 \mu'_1(v^{(k)})&=& k \kappa_1^{(\gamma_i)}\nonumber\\
 \mu'_2(v^{(k)})&=& k\kappa_2^{(\gamma_i)}+k^2(\kappa_1^{(\gamma_i)})^2\nonumber\\
 \mu'_3(v^{(k)})&=& k\kappa_3^{(\gamma_i)}+3k^2\kappa_2^{(\gamma_i)}\kappa_1^{(\gamma_i)} +k^3(\kappa_1^{(\gamma_i)})^3\nonumber\\
 \mu'_4(v^{(k)})&=& k\kappa_4^{(\gamma_i)}+4k^2\kappa_3^{(\gamma_i)}\kappa_1^{(\gamma_i)} +3k^2(\kappa_2^{(\gamma_i)})^2\nonumber\\
 &&+6 k^3\kappa_2^{(\gamma_i)}(\kappa_1^{(\gamma_i)})^2+k^4(\kappa_1^{(\gamma_i)})^4\nonumber\\
 \mu'_5(v^{(k)})&=& k\kappa_5^{(\gamma_i)}+5k^2\kappa_4^{(\gamma_i)}\kappa_1^{(\gamma_i)}
 +10k^2\kappa_3^{(\gamma_i)}\kappa_2^{(\gamma_i)}\nonumber\\
 &&+10k^3\kappa_3^{(\gamma_i)}(\kappa_1^{(\gamma_i)})^2 +15k^3(\kappa_2^{(\gamma_i)})^2\kappa_1^{(\gamma_i)}\nonumber\\
 &&+10k^4\kappa_2^{(\gamma_i)}(\kappa_1^{(\gamma_i)})^3
 +k^5(\kappa_1^{(\gamma_i)})^5\nonumber\\
 \mu'_6(v^{(k)})&=&\ldots\nonumber\ ,
\end{eqnarray}
in which $k$ is the number of detected photons, each one converted with its own $\gamma_i$ ($i=1,2,\ldots,k$), and the cumulants are those of the probability distribution $p_\gamma$. We observe that each term contains a product of cumulants in which the sum of the indices is the order of the moment.

Let us assume a narrow $p_\gamma$ distribution so that $\sigma^2/\bar{\gamma}^2 \rightarrow 0$. In terms of cumulants this rewrites $\kappa_2^{(\gamma_i)}= o[(\kappa_1^{(\gamma_i)})^2]$. Under this hypothesis also $\tilde{\mu}_3/\bar{\gamma}^3 \rightarrow 0$, that is $\kappa_3^{(\gamma_i)}= o[(\kappa_1^{(\gamma_i)})^3]$, as we can write $\tilde{\mu}_3/\bar{\gamma}^3= (\tilde{\mu}_3/\sigma^3)(\sigma^3/\bar{\gamma}^3)$, where the first factor is the (finite) coefficient of skewness of the distribution $p_\gamma$. Actually it can be easily shown that $\kappa_j^{(\gamma_i)}= o[(\kappa_1^{(\gamma_i)})^j]$ relations hold for any $j\ge 2$, if $\kappa_2^{(\gamma_i)}= o[(\kappa_1^{(\gamma_i)})^2]$. Taking into account that the number of detected photons, $k$, is a finite number, all the monomials in the above expressions of the moments are negligible with respect to the last one, so that we can approximate $\mu'_s(v^{(k)})\cong k^s (\kappa_1^{(\gamma_i)})^s= k^s \bar{\gamma}^s$. By substituting in Eq.~(\ref{eq:muV2}) we get
\begin{eqnarray}
 \mu_r(v)&=&\sum_{j=0}^{r} \left(
 \begin{array}{c}r\\j\end{array}\right)
 (-\langle m\rangle \bar{\gamma})^{r-j}
 \sum_{k=0}^{+\infty} P_{m=k} k^j \bar{\gamma}^j\nonumber\\
 &=& \bar{\gamma}^r \sum_{k=0}^{+\infty} P_{m=k}\sum_{j=0}^{r} \left(
 \begin{array}{c}r\\j\end{array}\right)
 k^j(-\langle m\rangle)^{r-j}\nonumber\\
 &=& \bar{\gamma}^r \sum_{k=0}^{+\infty} P_{m=k}
 (k-\langle m\rangle)^r= \bar{\gamma}^r \mu_r(m)\ .\label{eq:muV3}
\end{eqnarray}
Note that, as $\mu_r(m)$ never vanishes, even in the case of light in a single-mode Fock state because of the non unit quantum efficiency of the detectors, actually Eq.~(\ref{eq:muV3}) holds for measurements performed on optical fields with any statistics.

Dividing both members of Eq.~(\ref{eq:muV3}) by $\langle v\rangle$ yields
\begin{eqnarray}
 \frac{\mu_r(v)}{\langle v\rangle} = \bar{\gamma}^{r-1} \frac{\mu_r(m)}{\langle m\rangle}
 \ ,\label{eq:muVfinal}
\end{eqnarray}
while the exact results for $r=2$ and $r=3$ would be
\begin{eqnarray}
 \frac{\mu_2(v)}{\langle v\rangle} &=& \bar{\gamma} \left[\frac{\mu_2(m)}{\langle m\rangle}+\frac{\sigma^2}{\bar{\gamma}^2}\right]\label{eq:mu2final}\\
 \frac{\mu_3(v)}{\langle v\rangle} &=& \bar{\gamma}^2 \left[\frac{\mu_3(m)}{\langle m\rangle}+3\frac{\mu_2(m)}{\langle m\rangle}\frac{\sigma^2}{\bar{\gamma}^2} +\frac{\tilde{\mu}_3}{\bar{\gamma}^3}\right]\ ,\label{eq:mu3final}
\end{eqnarray}
respectively.

We thus assess that, when detector and processing electronics ensure a sufficiently small ratio $\sigma/\bar{\gamma}$, the scaling law in Eq.~(\ref{eq:muVfinal}) holds and the simple knowledge of $\bar{\gamma}$ allows reconstructing $P_m$. In fact binning the $v$ data of a measurement into bins of width $\bar{\gamma}$ produces a distribution $P_v$ identical to $P_m$. Alternatively we can say that $P_m$ is recovered by dividing the $v$ data by $\bar{\gamma}$ and then rebinning the new values into unitary bins.

How to determine $\bar{\gamma}$ when $\sigma\ll \bar{\gamma}$ has been already shown \cite{noiJMO}. Here we demonstrate that, for a detector simply endowed with linear response, we can both determine  $\bar{\gamma}$ and decide on the negligibility of $\sigma$ with respect to $\bar{\gamma}$.

At this point, for the first time in this work, we make use of Eq. ~(\ref{eq:phel}). Such a link between the $P_m$ and $P_n$ distributions gives $\langle m \rangle = \eta \langle n \rangle$  and  $\langle m^2 \rangle = \eta^2 \langle n^2 \rangle + \eta(1-\eta) \langle n \rangle$. Thus we find \cite{noiJMO}
\begin{eqnarray}
 \frac{\mu_2(m)}{\langle m\rangle} =\eta Q +1\ .\label{eq:fanoM}
\end{eqnarray}
where $Q=[\mu_2(n)-\langle n\rangle]/\langle n\rangle$ is the Mandel parameter of the light entering the experimental apparatus in Fig.~\ref{fig:1} and containing $\langle n \rangle$  photons in the $T_{\mathrm{M}}$ time interval \cite{MandelandWolf}. Substituting Eq.~(\ref{eq:fanoM}) into Eq.~(\ref{eq:muVfinal}) and taking into account Eq.~(\ref{eq:vAVER}) yield
\begin{eqnarray}
 \frac{\mu_2(v)}{\langle v\rangle} = \frac{Q}{\langle n\rangle}\langle v\rangle+\bar{\gamma}
 \left(1+\frac{\sigma^2}{\bar{\gamma}^2}\right)\ .\label{eq:fanoV}
\end{eqnarray}
in which $Q/{\langle n\rangle}$ is independent of $\eta$. On the other hand, $\langle v\rangle$ depends on $\eta$, which can be changed by acting on the light delivery optics: $\eta$ can be set at any value between the product of $\eta_\mathrm{q}$  times the coupling efficiency of the optical delivery system and zero by adding filter into the system that delivers the light to the detector. Thus by repeated measurements of the same light at different $\eta$, we can verify the linear dependence on $\langle v \rangle$ in Eq.~(\ref{eq:fanoV}). The experimental $\mu_2(v)/\langle v\rangle$ data plotted as a function of $\langle v\rangle$ should align along a straight line, whose intercept reduces to $\bar{\gamma}$ if $\sigma^2/\bar{\gamma}^2\ll 1$. Knowing $\bar{\gamma}$ allows proceeding to the rebinning of the $v$ data that leads to the reconstruction of $P_m$. Experimental applications to some non-trivial classical states are described in references \cite{noiJMO,ASL,WignerOL1}.

\section {Discussion}
The assessment that an experimental apparatus has a $\sigma/\bar{\gamma}$ ratio sufficiently small for the validity of Eq.~(\ref{eq:muVfinal}) deserves some comments, owing to the difficulty of knowing $\bar{\gamma}$ and $\sigma$ separately.

We first observe that for any photo-emissive detector $\sigma$ decreases at increasing the strength of the electric field experienced by the photoelectrons as soon as they leave the cathode. For a PMT in which the internal gain is provided by multi-dynode cascade amplification, increasing the voltage difference between cathode and first dynode produces smaller $\sigma$ values. For PMT's in which the electron amplification is provided by other structures (\textit{e.g.} micro-channel-plate, metal channels, etc.), the same effect is obtained by acting on the voltage of the accelerating electrode. For a HPD, in which the electrons released by the photocathode are multiplied by a reverse biased avalanche diode, $\sigma$ is lowered by applying greater negative high voltages to the photocathode.

In the case of PMT's, modifying the voltage partition to change $\sigma$ brings about a change in $\bar{\gamma}$ that cannot be easily compensated by acting on the overall voltage applied between anode and cathode. In the case of HPD's this compensation is feasible by adjusting the avalanche diode reverse bias voltage. However in any electronic apparatus that processes the detector output there is a step that allows changing $\bar{\gamma}$ by a known factor (\textit{e.g.} in Fig.~\ref{fig:1}: both AMPL gain and ADC scale) while keeping $\sigma/\bar{\gamma}$  virtually constant.

The expression of the intercept in Eq.~(\ref{eq:fanoV}) is such that, upon changing $\bar{\gamma}$ by a known factor but not $\sigma/\bar{\gamma}$, a new series of measurements at different $\eta$ values would provide a new evaluation of the intercept, whose value should scale by the same factor. On the contrary, for constant $\bar{\gamma}$ and different $\sigma/\bar{\gamma}$ ratios, the intercept should change differently. Note that a check of the constancy of $\bar{\gamma}$ is provided by Eq.~(\ref{eq:vAVER}) in which $\sigma$ does not appear. If, by manipulating the voltages supplied to PMT/HPD detectors as described to change $\sigma$, we achieve a situation of constant and minimum intercept, we have proved that $\sigma^2/\bar{\gamma}^2\ll 1$ in Eq.~(\ref{eq:fanoV}). We can thus use this limit value of the intercept as the correct $\bar{\gamma}$ to rebin the experimental $P_v$ distribution and reconstruct $P_m$. If the $\langle m\rangle$ value provided by the reconstructed $P_m$ fits Eq.~(\ref{eq:vAVER}), it means that the detector guarantees a $\sigma^2/\bar{\gamma}^2$ not only much smaller than one but small enough for the validity of Eq.~(\ref{eq:muVfinal}).

We finally note that the above described verifications of the validity of Eq.~(\ref{eq:muVfinal}) are self-consistent in that they do not require measuring $p_\gamma$ to establish the relation between $\sigma^2$ and $\bar{\gamma}^2$. This is a definite advantage with respect to any potential method for $P_m$ reconstruction based on the determination of the $p_\gamma$ moments \cite{IEEE}.

\section{Conclusions}
We have shown that for any linear detector we can both measure $\bar{\gamma}$ and determine if $\sigma/\bar{\gamma}$ is sufficiently small for taking as reliable the $P_m$ reconstruction achieved by binning the experimental $v$ values into bins of width $\bar{\gamma}$. For the method to work it is necessary that the $m$-range where $P_m$ is non-negligible falls within the linearity range of the apparatus, which must be broad enough for a satisfactory verification of Eq.~(\ref{eq:fanoV}). In forthcoming papers we will show that our method works not only with HPD's \cite{noiJMO,ASL} and the Burle 8850 PMT \cite{noiRSI,noiJMO}, but also with detectors such as Si multi-pixel photon detectors \cite{Israel}. and more PMT's endowed with single photon sensitivity. Useful detectors might also be solid state detectors such as avalanche photodiodes in the linear amplification regime \cite{APDlowGAIN}. By the way, some photon-number resolution is being demonstrated for these detectors, in particular if connected to charge-integration readout circuits with sufficiently low noise \cite{refLINK}. At last, for a thermal detector such as a TES, obtaining a $\sigma$ value sufficiently smaller than $\bar{\gamma}$ would be a minimal performance as compared to the excellent photon-number resolving power demonstrated by Lita \emph{et al.} \cite{Lita} up to 7 detected photons and might allow using a less sophisticated apparatus.

We think that the results described in this paper will broaden the choice of detectors suitable for measuring photon statistics. The essential requirement for the detector, beside that of the linearity of the response, is the smallness of the ratio $\sigma^2/\bar{\gamma}^2$, which can be ascertained (see above) without measuring $p_\gamma$.
The fact that the method applies to measurements in the macroscopic realm may turn out to be relevant in all cases in which one cannot attenuate the light to bring the photon detection rate down to the regime where photon-counters operate. As examples we mention fields produced by events either rare or unstable and, more importantly, all nonclassical fields, where our method risks being the only one applicable to macroscopic fields.
\subsection*{Acknowledgments}
The authors are indebted to E. Casini (Como) for his useful suggestions and to A. Allevi (C.N.I.S.M., U.d.R. Milano) for her collaboration to the experimental work that has stimulated the present study.



\end{document}